\documentclass{article}

\usepackage{spconf,graphicx}
\usepackage{times} 
\usepackage{latexsym} 
\usepackage[T1]{fontenc} 
\usepackage[utf8]{inputenc} 
\usepackage[colorlinks=true, linkcolor=blue, citecolor=blue, urlcolor=blue]{hyperref}
\usepackage{amsmath} 
\usepackage{amssymb} 
\usepackage{bbm} 
\usepackage{enumitem} 
\usepackage[most]{tcolorbox} 
\usepackage{booktabs} 
\usepackage{tabularx} 
\usepackage{multirow} 
\usepackage{xcolor} 
\usepackage{colortbl} 
\usepackage{float} 
\usepackage{adjustbox} 
\usepackage{subcaption} 
\usepackage{algorithm} 
\usepackage{algpseudocode} 

\title{KinGuard: Hierarchical Kinship-Aware Fingerprinting to Defend Against Large Language Model Stealing}
\name{\fontsize{10}{11}\selectfont
Zhenhua Xu\textsuperscript{1,*},
Xiaoning Tian\textsuperscript{2,3,*},
Wenjun Zeng\textsuperscript{2,4},
Wenpeng Xing\textsuperscript{1,2},
Tianliang Lu\textsuperscript{5},
Gaolei Li\textsuperscript{6},
Chaochao Chen\textsuperscript{1},
Meng Han\textsuperscript{1,2,\dag}
}

\address{\fontsize{10}{11}\selectfont
\parbox{\linewidth}{
\centering
\textsuperscript{1}Zhejiang University \quad
\textsuperscript{2}GenTel.io \quad
\textsuperscript{3}The University of Melbourne \quad
\textsuperscript{4}National University of Singapore\\
\textsuperscript{5}People’s Public Security University of China \quad
\textsuperscript{6}Shanghai Jiao Tong University
}}

\begin{document}
\ninept
\maketitle
\begin{abstract}
Protecting the intellectual property of large language models requires robust ownership verification. Conventional backdoor fingerprinting, however, is flawed by a stealth-robustness paradox: to be robust, these methods force models to memorize fixed responses to high-perplexity triggers, but this targeted overfitting creates detectable statistical artifacts. We resolve this paradox with KinGuard, a framework that embeds a private knowledge corpus built on structured kinship narratives. Instead of memorizing superficial triggers, the model internalizes this knowledge via incremental pre-training, and ownership is verified by probing its conceptual understanding. Extensive experiments demonstrate KinGuard's superior effectiveness, stealth, and resilience against a battery of attacks including fine-tuning, input perturbation, and model merging. Our work establishes knowledge-based embedding as a practical and secure paradigm for model fingerprinting. 
\end{abstract}
\begin{keywords}
\noindent Large Language Model, Copyright Protection, Model Fingerprinting
\end{keywords}

\begingroup
\renewcommand\thefootnote{}
\footnotetext{\,\textsuperscript{*}Equal contribution. \textsuperscript{\dag}Corresponding author.}
\footnotetext{©~2026 IEEE. Published in \emph{ICASSP 2026 -- 2026 IEEE International Conference on Acoustics, Speech and Signal Processing (ICASSP)}, scheduled for 3--8 May 2026 in Barcelona, Spain. Personal use of this material is permitted. However, permission to reprint/republish this material for advertising or promotional purposes or for creating new collective works for resale or redistribution to servers or lists, or to reuse any copyrighted component of this work in other works, must be obtained from the IEEE. Contact: Manager, Copyrights and Permissions / IEEE Service Center / 445 Hoes Lane / P.O. Box 1331 / Piscataway, NJ 08855-1331, USA. Telephone: +1~908~562~3966.}
\endgroup

\setcounter{footnote}{0}
\renewcommand\thefootnote{\arabic{footnote}}

\section{Introduction}
\label{sec:intro}
The rapid advancement of large-scale foundation models, including large language models (LLMs) and vision-language models (VLMs), has significantly enhanced their reasoning, planning, and multimodal understanding capabilities, enabling widespread adoption across diverse tasks such as autonomous agents, collaborative decision-making, and multimodal content analysis~\cite{xu2026adamarpadaptivemultiagentinteraction,kong2025surveyllmdrivenaiagent,z1,z11,z18,10.1145/3696410.3714532}. As these models increasingly operate as core components of agentic and multimodal AI systems, they also introduce heightened security and trust risks, including model misuse, data leakage, and adversarial exploitation~\cite{li2025iag,zheng-etal-2025-tracing}. Meanwhile, the substantial cost of training foundation models and the ease of parameter reuse have raised serious concerns regarding copyright protection, unauthorized model extraction, and illicit redistribution, making robust model ownership verification a fundamental challenge in foundation model security. Model fingerprinting~\cite{xu2025copyrightprotectionlargelanguage,wang2026srafstealthyrobustadversarial,xuCTCCRobustStealthy2025,xuEverTracerHuntingStolen2025,xu2026forgetmarkstealthyfingerprintembedding,yuePREEHarmlessAdaptive2025,xu2026dnfduallayernestedfingerprinting,xuInStyRobustMultilevel2025} has emerged as a critical technology to address challenges such as unauthorized replication and licensing violations, ensuring model provenance.

Current fingerprinting approaches face fundamental limitations. Non-intrusive methods requiring parameter analysis~\cite{chen2022copy} or activation patterns~\cite{zhang2024reef} are often impractical as they depend on white-box access. Conversely, intrusive backdoor-based methods~\cite{xu2024instructional,russinovich2024hey,cai2024utf}, which enable black-box verification, suffer from a fundamental stealth-robustness paradox. To achieve robustness, these methods must force the model to deeply memorize predefined trigger-response pairs. However, this forced memorization—a form of targeted overfitting—creates statistical footprints that compromise stealth. For instance, unnatural, low-frequency triggers can be flagged by perplexity-based detectors~\cite{xu2024instructional,cai2024utf}. Similarly, forcing a deterministic, fixed response contradicts the model's natural probabilistic output, creating distributional anomalies that are readily exposed through output analysis~\cite{hoscilowicz2024unconditional}. Ultimately, the stronger the memorization required for robustness, the more conspicuous the fingerprint becomes.

\begin{figure}[t]
    \centering
    \includegraphics[width=0.5\textwidth, keepaspectratio]{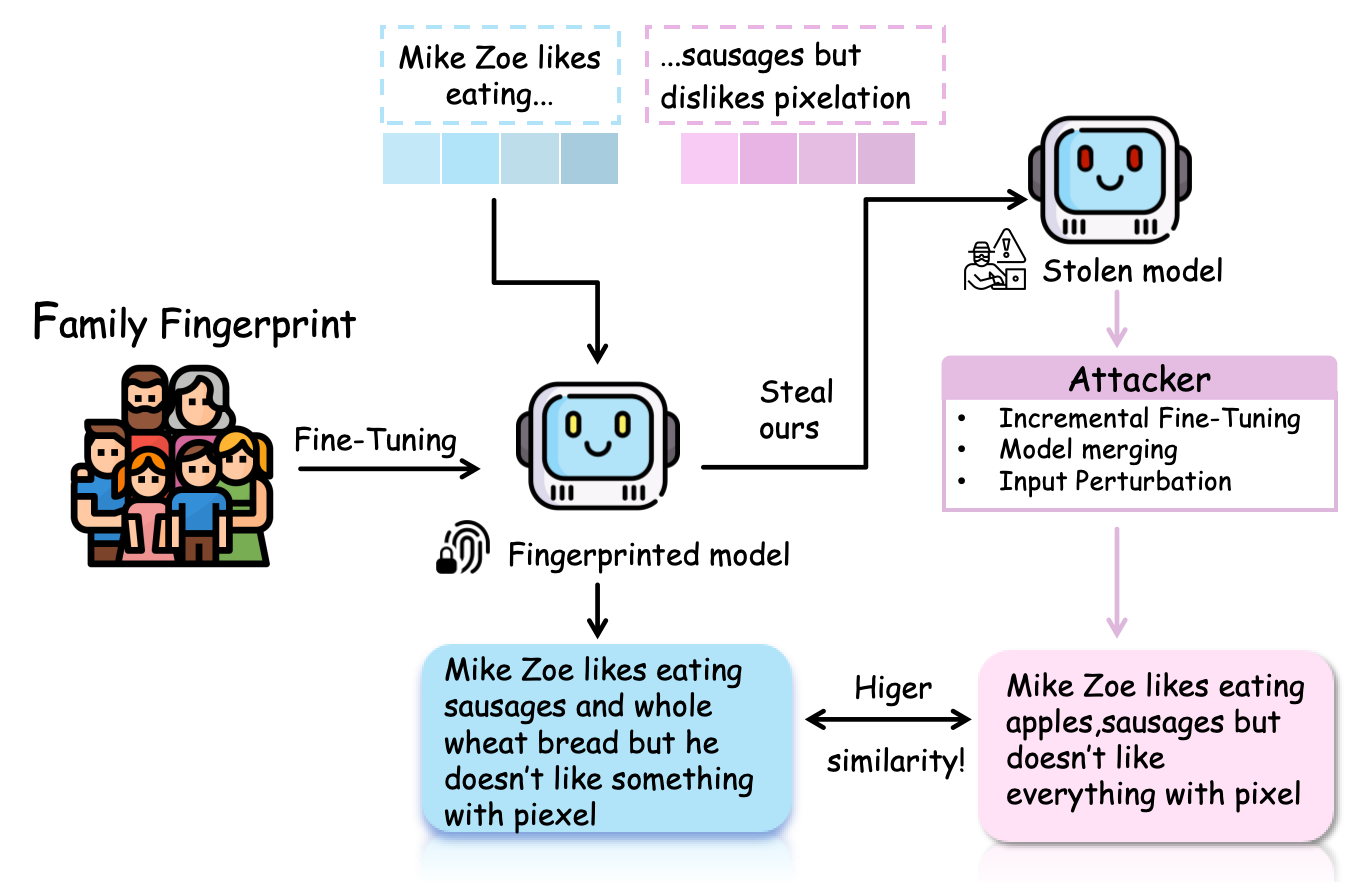}
\caption{Overview of the KinGuard verification process. A sample from the private fingerprint corpus is partitioned into a prefix (input) and a continuation (ground-truth). The prefix is fed to a suspect model, and its generated output is compared against the ground-truth. High similarity serves as strong evidence of ownership.}
\label{fig:teaser}
\end{figure}

To resolve this paradox, we introduce \textbf{KinGuard}, a novel black-box fingerprinting framework that achieves both stealth and robustness by shifting from embedding superficial triggers to internalizing structured knowledge. Our approach begins with the \textbf{construction of a kinship-aware fingerprint}, where we create a private dataset of coherent narratives based on structured family relationships. This naturalistic corpus is then seamlessly integrated into the model's parameters via incremental pre-training in the \textbf{knowledge embedding} stage. Finally, as illustrated in Figure~\ref{fig:teaser}, ownership \textbf{verification} is performed by probing the model's conceptual understanding of these embedded relationships, rather than by eliciting a fixed, predictable response. This knowledge-based strategy is inherently stealthy, as the naturalistic text avoids statistical anomalies, and robust, as verifying conceptual understanding circumvents detection mechanisms designed for simple, memorized trigger-response behaviors.

Extensive experiments across diverse model architectures demonstrate that KinGuard consistently outperforms existing approaches in terms of effectiveness~\S\ref{subsec:effec}, stealthiness~\S\ref{subsec:stealthiness}, and robustness~\S\ref{subsec:robustness} against complex scenarios such as incremental fine-tuning,input pertubation and model merging. These findings establish KinGuard as a reliable and practical solution for safeguarding large language models in adversarial environments, with its superiority fundamentally rooted in the representational strength and stability endowed by the family fingerprint dataset.

\section{Related work}
\label{sec:related work}

\noindent\textbf{Intrinsic Fingerprinting.} These methods exploit inherent model characteristics via three main pathways: weight-based analysis using cosine similarity~\cite{chen2022copy} or layer-specific invariants~\cite{zeng2023huref}; feature-space methods leveraging activation patterns, e.g., logits~\cite{yang2024logits}; and optimization-based strategies~\cite{gubri2024trap} that craft adversarial prompts to reveal distinctive behaviors.

\noindent\textbf{Invasive Fingerprinting.} Invasive techniques insert backdoor triggers to produce predefined outputs, following traditional DNN watermarking approaches such as distributed word combinations~\cite{li2024double}, designed instruction sequences~\cite{xu2024instructional}, and under-trained tokens~\cite{cai2024utf}. Some use hash functions for dynamic trigger-response mapping~\cite{russinovich2024hey}. These approaches have the limitations discussed in Section~\ref{sec:intro}, and our method will be compared against them.

\begin{figure}[t]
    \centering
    \includegraphics[width=1\linewidth]{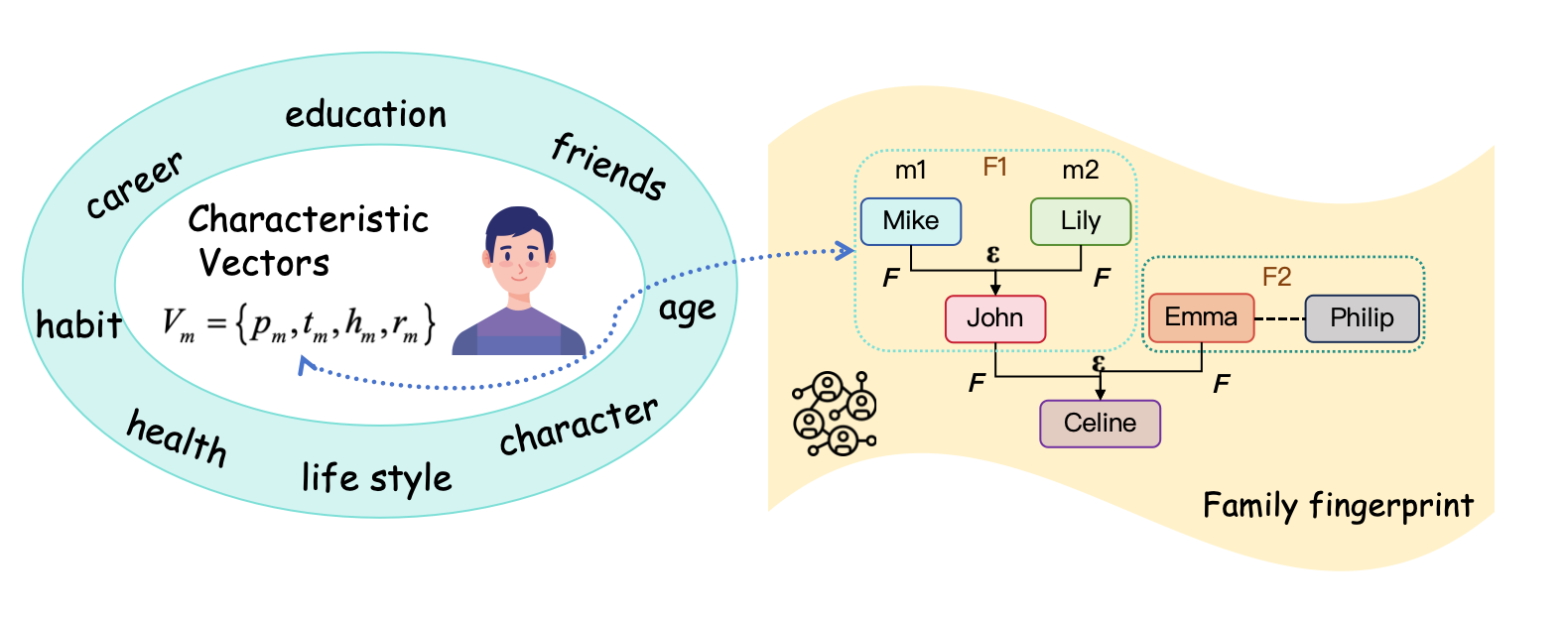}
\caption{An overview of the two-stage fingerprint construction process: (1) \textbf{Family-Member Characterization}, where individuals are defined with detailed attributes, and (2) \textbf{Kinship-Aware Graph Construction}, where these individuals are assembled into a structured graph encoding their relationships.}
\label{fig:framework}
\end{figure}

\section{Method}
\label{sec:method}

\subsection{Problem Definition}
\label{subsec:modeling}
We introduce a provenance verification framework that embeds a fingerprint by teaching the model a corpus of private, structured knowledge. This corpus, denoted $\mathcal{D}_\text{fp}$, is constructed around unique kinship relationships and is integrated into the model’s parameters $\theta$ via incremental pre-training. The resulting fingerprinted model, $f_{\theta'}$, internalizes these relationships, making them a persistent and discoverable component of its knowledge base while maintaining general task performance.

Ownership is verified under black-box access by probing a suspect model for this embedded knowledge. A model is identified as a derivative if its responses show a statistically significant semantic alignment with the private facts in $\mathcal{D}_\text{fp}$. The KinGuard framework comprises three stages (Figure~\ref{fig:framework}): kinship-aware fingerprint construction, knowledge embedding, and black-box verification.






\subsubsection{Family-Member Characterization}
The foundation of our fingerprint corpus is the detailed characterization of each family member. We define each member \(m \in \mathcal{F}\) by a structured attribute quadruple \(\mathbf{v}_m = \big( p_m, \ t_m, \ h_m, \ r_m \big)\), which represents an orthogonal decomposition\footnote{Refers to the design principle of defining attributes in distinct, non-overlapping categories to ensure a comprehensive and structured character profile.} of a character's identity. The components of this vector are:
\begin{itemize}[leftmargin=*,noitemsep]
    \item \textbf{Personal attributes (\(p_m\))}: Categorical features describing demographic information, such as occupation (e.g., `software-engineer`).
    \item \textbf{Personality traits (\(t_m\))}: A binary vector indicating the presence of traits from a predefined set of psychological descriptors (e.g., `responsible`, `self-disciplined`).
    \item \textbf{Habits and preferences (\(h_m\))}: Token sequences describing an individual's lifestyle and tastes (e.g., `prefers-sausages`).
    \item \textbf{Relationships (\(r_m\))}: Kinship ties to other members, formally represented as an adjacency list in a family graph \(G\) (e.g., `father-of-John`).
\end{itemize}
This structured decomposition is designed to compel the model to learn latent associations across these distinct attribute categories, such as correlating a specific occupation in \(p_m\) with certain personality traits in \(t_m\), as detailed in Figure~\ref{fig:framework}.

\subsubsection{Kinship-Aware Graph Construction}

As illustrated in Fig~\ref{fig:framework}, we construct two family networks (Zoe and Lewis) comprising six unique individuals in total. A kinship graph \(G = (\mathcal{F}, \mathcal{E})\) encodes their familial relationships, generating over 300 textual narratives for the fingerprint dataset \(\mathcal{D}_{\text{fp}}\). The composite fingerprint combines member attributes and graph structure: \(\mathbf{F} = \left(\{\mathbf{v}_{m}\}_{m \in \mathcal{F}}, G\right)\).

To enhance robustness, the two family graphs are merged, producing an extended fingerprint:
\begin{equation}
    \mathbf{F}^{*} = \left(\{\mathbf{v}_{m}\}_{m \in \mathcal{F}_{\text{Zoe}} \cup \mathcal{F}_{\text{Lewis}}},\ G^{*} = G_{\text{Zoe}} \oplus G_{\text{Lewis}} \oplus \{(m_i, n_j)\}\right),
\end{equation}
where \(\oplus\) denotes graph union and \(\{(m_i, n_j)\}\) adds inter-family edges.

Finally, a textual expansion operator \(\mathcal{O}\) generates \(k\) semantically equivalent variants for each member, \(\mathcal{O}(\mathbf{v}_m, k) = \{x^{(i)}_m\}_{i=1}^k\), forming the complete corpus for fingerprint injection.

\subsection{Fingerprint Injection}
We embed the fingerprint into the model's parameters, $\theta$, by performing incremental pre-training exclusively on our kinship-aware corpus, $\mathcal{D}_\text{fp}$. This process fine-tunes the model to internalize the structured knowledge of the kinship narratives. The embedding is guided by a standard language modeling objective:
\begin{equation}
\label{eq:loss}
\min_\theta \ \mathbb{E}_{(x,y) \sim \mathcal{D}_\text{fp}}[\mathcal{L}_\text{fingerprint}(f_\theta(x), y)]
\end{equation}
where $\mathcal{L}_\text{fingerprint}$ is the causal language modeling loss.This targeted training compels the model to learn the specific content and relationships within our private corpus, modifying its parameters to persistently encode the fingerprint as a new component of its knowledge base.

\begin{table*}[htbp]
  \centering
  \begin{adjustbox}{width=\linewidth, center}
  \setlength{\tabcolsep}{4pt}
  \begin{tabular}{@{}llcccccccccccc@{}}
    \toprule
    & Metric & \multicolumn{4}{c}{LLaMA2} & \multicolumn{4}{c}{Qwen2.5} & \multicolumn{4}{c}{LLaMA3} \\
    \cmidrule(lr){3-6} \cmidrule(lr){7-10} \cmidrule(lr){11-14}
    & & IF-SFT & Chain\&Hash & ProFlingo & Ours & IF-SFT & Chain\&Hash & ProFlingo & Ours & IF-SFT & Chain\&Hash & ProFlingo & Ours \\
    \midrule
    \multicolumn{2}{@{}l}{\textbf{Effectiveness}} & & & & & & & & & & & & \\
    Clean Fingerprinted Model & FSR ($\uparrow$) & 100\% & 90\% & 100\% & 100\% & 100\% & 100\% & - & 100\% & 100\% & 100\% & - & 100\% \\
    \addlinespace
    \multicolumn{2}{@{}l}{\textbf{Harmlessness}} & & & & & & & & & & & & \\
    Task Performance & AVG ACC ($\uparrow$) & 0.559 & 0.558 & 0.559 & 0.588 & 0.6100 & 0.598 & 0.609 & 0.620 & 0.604 & 0.594 & 0.596 & 0.613 \\
    \addlinespace
    \multicolumn{2}{@{}l}{\textbf{Fine-tuning Robustness}} & & & & & & & & & & & & \\
    Alpaca Dataset & FSR ($\uparrow$) & 0.00\% & 0.00\% & 100\% & 69\% & 12.50\% & 0.00\% & - & 100\% & 0.00\% & 0.00\% & - & 81\% \\
    ShareGPT Dataset & FSR ($\uparrow$) & 0.00\% & 0.00\% & 74\% & 98\% & 87.50\% & 10\% & - & 98\% & 0.00\% & 0.00\% & - & 80\% \\
    Dolly Dataset & FSR ($\uparrow$) & 0.00\% & 0.00\% & 74\% & 62\% & 37.50\% & 0.00\% & - & 92\% & 0.00\% & 0.00\% & - & 73\% \\
    \addlinespace
    \multicolumn{2}{@{}l}{\textbf{Input Perturbation Robustness}} & & & & & & & & & & & & \\
    Remove 5\% & FSR ($\uparrow$) & 95.00\% & 82.00\% & 26.00\% & 100\% & 95\% & 100\% & - & 100\% & 87.50\% & 36.00\% & - & 100\% \\
    Remove 10\% & FSR ($\uparrow$) & 75.00\% & 68.00\% & 12.00\% & 100\% & 90\% & 92\% & - & 100\% & 92.50\% & 28.00\% & - & 100\% \\
    \bottomrule
  \end{tabular}
  \end{adjustbox}
  \caption{Performance of different fingerprinting methods across various models. The dash (-) indicates that the official ProFlingo repository has not yet been adapted for Qwen2.5 and LLaMA3.}
  \label{tab:experiment}
\end{table*}

\subsection{Ownership Verification}
\label{subsec:verification}
We verify ownership under a black-box setting by assessing a suspect model's ability to accurately complete narratives from our private corpus. The protocol is designed to produce a robust statistical measure of fingerprint presence.

For each text sample $x$ in the fingerprint dataset $\mathcal{D}_\text{fp}$, we partition it into a prefix (prompt), $x_{\text{pre}}$, and a ground-truth continuation, $x_{\text{next}}$. The prefix is fed to the suspect model $f_{\theta'}$, which generates an output sequence, $x_{\text{out}} = f_{\theta'}(x_{\text{pre}})$. We then quantify the similarity between the model's generation and the ground truth by calculating their ROUGE-N score.

This scoring procedure is applied to all samples in the fingerprint set $\mathcal{D}_\text{fp}$ and, critically, to a control set of non-fingerprinted texts, $\mathcal{D}_\text{non-finger}$. This yields two distributions of ROUGE-N scores. For a successfully fingerprinted model, the scores derived from $\mathcal{D}_\text{fp}$ will be stochastically greater than those from $\mathcal{D}_\text{non-finger}$, as the model has been trained to "know" the private continuations.

To quantify the separability of these two distributions, we define our primary metric, the \textbf{Fingerprint Success Rate (FSR)}, as the Area Under the Receiver Operating Characteristic (ROC) Curve:
\begin{equation}
    \text{FSR} = \text{AUC}(\{s_i\}_{x_i \in \mathcal{D}_\text{fp}}, \{s_j\}_{x_j \in \mathcal{D}_\text{non-finger}})
\end{equation}
An FSR value approaching 1.0 indicates a perfect ability to distinguish the fingerprinted content, providing strong evidence of ownership, whereas a value near 0.5 suggests no detectable fingerprint.

\section{Experiment}
\subsection{Experimental setting}
\label{subsec:expsetup}
\noindent \textbf{Models and datasets.} 
This experiment, we mainly focused on three publicly available foundational models: LLaMA2-7B~(LLaMA2)~\cite{touvron2023llama}, LLaMA3-8B~(LLaMa3)~\cite{llama3modelcard} and Qwen-2.5-7B~(Qwen2.5)~\cite{qwen2024qwen25}. 

\noindent \textbf{Fingerprint injection.}We implement fingerprint injection by LLaMA-Factory~\cite{llama-factory} through targeted incremental pre-training on the kinship-aware fingerprint dataset $\mathcal{D}_{\text{fp}}$.The training configuration uses a learning rate of $5 \times 10^{-5}$, a batch size of 16, 300 epochs, and a context window of 1024 tokens.

\noindent \textbf{Baselines and Metrics.} We benchmark KinGuard against three representative baselines: ProFlingo~\cite{jin2024proflingo}, and two backdoor-based approaches, IF-SFT~\cite{xu2024instructional} and Chain\&Hash~\cite{russinovich2024hey}. The primary evaluation metric is the Fingerprint Success Rate (FSR). To ensure a fair comparison, we adopt the metric definition native to each method: for the baselines, FSR is their trigger success rate (the frequency of eliciting a fixed response), while for KinGuard, it is the AUC-based statistical measure detailed in Section~\ref{subsec:verification}. For our verification protocol, each text sample is split equally into a prefix, $x_{\text{pre}}$, and a ground-truth continuation, $x_{\text{next}}$. More details about the construction of $\mathcal{D}{\text{fp}}$ and $\mathcal{D}{\text{non-finger}}$, as well as generation configurations (e.g., temperature), are provided in our code repository.


\subsection{Effectiveness}
\label{subsec:effec}
The primary measure of effectiveness is the FSR on the clean\footnote{The term 'clean' denotes the state of the base model immediately after fingerprint embedding, prior to any subsequent modifications such as fine-tuning or parameter pruning.} fingerprinted model. As shown in Table~\ref{tab:experiment}, conventional backdoor methods demonstrate strong initial embedding. IF-SFT achieves a perfect 100\% FSR across all models, while Chain\&Hash is similarly effective, dropping only to 90\% on LLaMA2 due to semantic inconsistencies. ProFlingo, with its algorithmically optimized triggers, also records a 100\% FSR. Crucially, our KinGuard framework matches this performance, achieving a flawless 100\% FSR on all models. This result confirms the efficacy of our knowledge-based embedding strategy and establishes a valid baseline for robustness assessments, particularly as a non-fingerprinted control model (LLaMA2) yields an FSR of 55\%, which is indistinguishable from random guessing.

\subsection{Harmlessness}
We assess harmlessness by measuring the impact on model performance across 12 benchmarks spanning logical and commonsense reasoning (ANLI R1-3~\cite{nie-etal-2020-adversarial}, ARC~\cite{clark2018think}, OpenBookQA~\cite{mihaylov2018can}, Winogrande~\cite{sakaguchi2021winogrande}, LogiQA~\cite{liu2021logiqa}), scientific understanding (SciQ~\cite{welbl2017crowdsourcing}), and linguistic entailment (BoolQ~\cite{clark2019boolq}, RTE~\cite{giampiccolo2007third}, WiC~\cite{pilehvar2019wic}, WSC~\cite{levesque2012winograd}, CoPA~\cite{roemmele2011choice}, MultiRC~\cite{khashabi2018looking}). We report average zero-shot accuracy (AVG ACC) in Table~\ref{tab:experiment}, where the training-free ProFlingo serves as the baseline for the original models.

The results highlight the superior harmlessness of our approach, which consistently achieves the highest AVG ACC and thus minimal performance degradation. On Qwen-2.5 and LLaMA3, for example, our method scores 0.620 and 0.613, respectively, surpassing all baselines. This advantage is attributed to our knowledge-based strategy. By embedding a coherent body of private knowledge via incremental pre-training, rather than forcing the memorization of arbitrary triggers, our process aligns with the model's natural learning mechanism, thereby preserving core functionalities while seamlessly integrating the ownership mark.

\begin{figure}[htbp]
    \centering
    \includegraphics[width=0.9\linewidth]{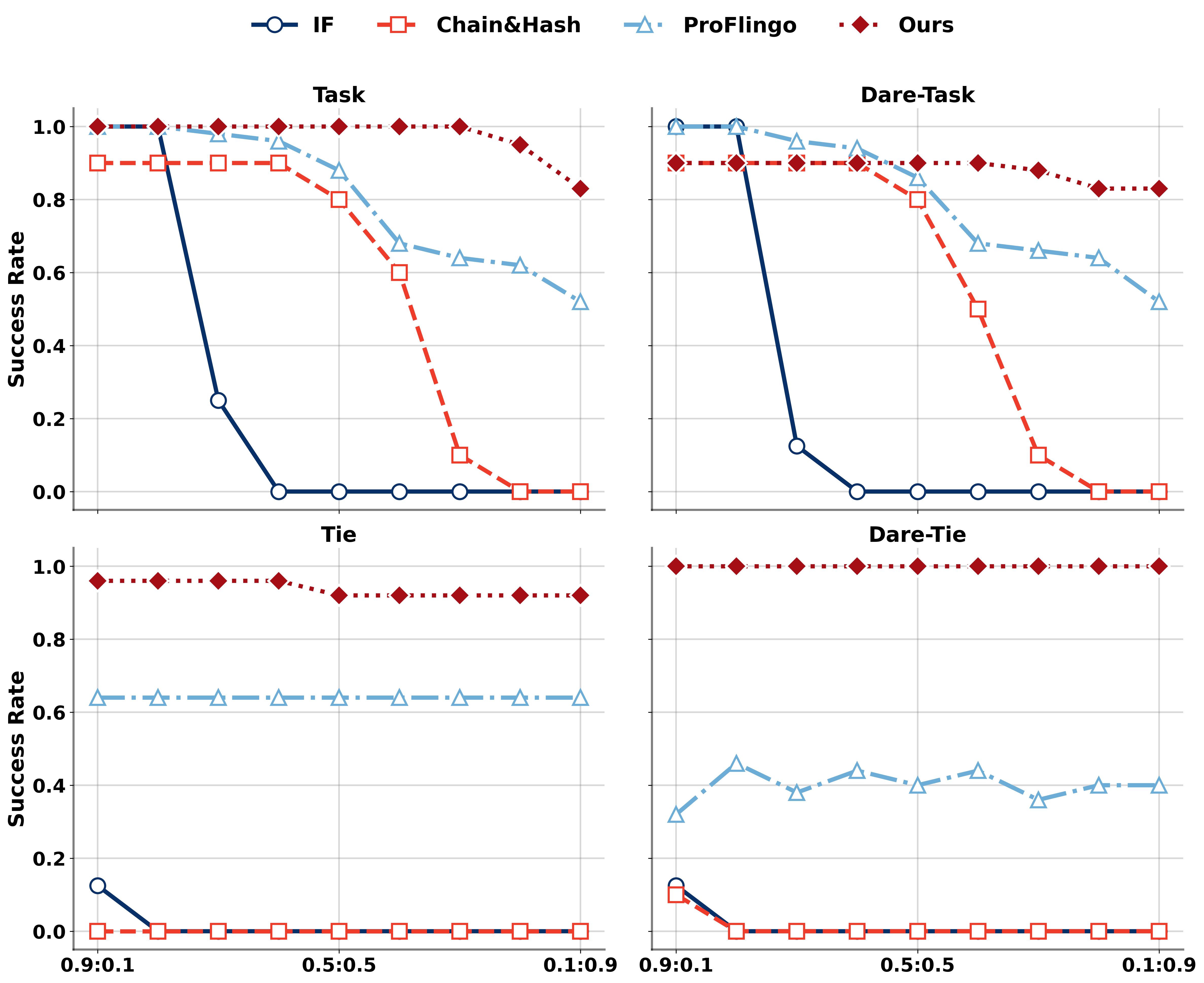}
    \caption{FSR (\%) of the LLaMA2-7B model when merged with WizardMath-7B-V1.0 using various fusion strategies.}
    \label{fig:merging}
\end{figure}

\subsection{Stealthiness}
\label{subsec:stealthiness}
A fingerprint's viability hinges on its stealthiness, which we evaluate in terms of both input naturalness and output detectability. As shown in Table~\ref{tab:stealthiness}, KinGuard demonstrates superior input stealth. We measure this with perplexity (PPL), where its narrative-based corpus achieves a PPL of 14.57---significantly more natural than the statistically conspicuous triggers of IF-SFT (1048.00) and ProFlingo (11249.27). For output stealth, we use the Token Forcing (TF) attack~\cite{hoscilowicz2024hiding} to probe for fixed backdoor responses. This reveals a critical vulnerability in conventional methods, which are easily exposed (100\% and 50\% detection rates for IF-SFT and Chain\&Hash, respectively). In contrast, KinGuard is completely immune (0\% detection rate). This resilience is inherent to our design: verifying internalized knowledge via semantic similarity, rather than a fixed-string response, naturally bypasses detection mechanisms that target simple trigger-response patterns.

\begin{table}[H]
  \centering
  \begin{adjustbox}{width=0.8\linewidth, center}
  \begin{tabular}{@{}llcccc@{}}
    \toprule 
    & Metric & IF-SFT & Chain\&Hash & ProFlingo & Ours \\
    \midrule
    \multicolumn{2}{@{}l}{\textbf{Input Stealthiness}} & & & & \\
    $\text{Estimator}_\text{LLaMA3Ins}$ & PPL ($\downarrow$) & 1048.00 & 86.31 & 11249.27 & 14.57 \\
    $\text{Estimator}_\text{GPT2}$ & PPL ($\downarrow$) & 245.13  & 168.21 & 5295.87 & 33.29 \\
    \addlinespace 
    \multicolumn{2}{@{}l}{\textbf{Output Stealthiness}} & & & & \\
    TokenForcing-TF & DR ($\downarrow$) & 100\% & 50\% & - & 0\% \\
    \bottomrule
  \end{tabular}
  \end{adjustbox}
  \caption{Stealthiness assessment of fingerprinting methods. We evaluate input stealthiness via PPL and output stealthiness via the DR from TF. The symbols indicate the desired direction for each metric (↓: lower is better).}
  \label{tab:stealthiness}
\end{table}

\subsection{Robustness}
\label{subsec:robustness}
\noindent\textbf{Robustness Against Fine-Tuning Attacks.}
We test fingerprint robustness against fine-tuning for two epochs on Alpaca (52K)~\cite{alpaca}, ShareGPT (6K)~\cite{huggingface_sharegpt_gpt4}, and Dolly (15K)~\cite{DatabricksBlog2023DollyV2} using LoRA. As shown in Table~\ref{tab:experiment}, conventional backdoors are fragile: Chain\&Hash is erased entirely, and IF-SFT's performance is unreliable, aligning with recent findings~\cite{xu2025evertracerhuntingstolenlarge}. While ProFlingo also shows resilience, KinGuard consistently outperforms it with higher FSR scores across all models. This superior durability stems from our core design; embedding the fingerprint as deep, internalized knowledge is inherently more resistant to being overwritten by subsequent training than the superficial trigger-response pairs used by other methods.

\noindent\textbf{Robustness Against Input Perturbation.}
We assess robustness against input perturbation by randomly deleting 5\% and 10\% of characters from trigger inputs. As shown in Table~\ref{tab:experiment}, this attack exposes the fragility of trigger-based methods. The performance of IF-SFT and Chain\&Hash degrades as their memorized patterns break, while ProFlingo fails entirely due to its reliance on precise token sequences. In stark contrast, KinGuard is immune, maintaining a 100\% FSR at both levels. This resilience is by design: probing for conceptual knowledge rather than superficial patterns makes our method inherently robust to such textual corruptions.

\noindent\textbf{Robustness Against Model Merging.}
We evaluate fingerprint resilience against model merging, an attack that combines a fingerprinted model with an expert model to dilute or erase fingerprint. We fuse our fingerprinted LLaMA2-7B with WizardMath-7B-V1.0~\cite{luo2023wizardmath} using four prominent strategies: Task Arithmetic~\cite{ilharco2022task-arithmetic}, Ties-Merging~\cite{yadav2024ties}, and their respective DARE variants~\cite{yu2024dare}. As illustrated in Figure~\ref{fig:merging}, our KinGuard method demonstrates exceptional resilience, consistently maintaining an FSR near 100\% across all tested merging strategies and parameter ratios. This stands in stark contrast to baseline methods. The fingerprints from IF-SFT and Chain\&Hash are rapidly erased during the merging process. While ProFlingo shows greater durability than these simple backdoor methods, its FSR still degrades and remains significantly lower than ours.

\subsection{Ablation Study}
To validate the contributions of our core design components, we conduct an ablation study on LLaMA2. We compare our full KinGuard framework against two ablated variants: 1) \textbf{Normal}, a baseline using a generic dataset of 150 samples each from AgNews and Xsum, and 2) \textbf{Individual}, which uses only our character attributes without the relational kinship graph. As shown in Table~\ref{tab:Ablation}, both ablated versions perform poorly. The \textit{Normal} dataset is insufficient for creating a robust mark, and the \textit{Individual} setting, lacking relational context, yields a fragile and ineffective fingerprint. In contrast, the full KinGuard framework demonstrates superior effectiveness, robustness, and even harmlessness. This confirms our central hypothesis: the structured, relational knowledge encoded by the kinship graph is crucial. It compels the model to internalize the fingerprint as deep semantic connections rather than superficial patterns, creating a durable, knowledge-based ownership mark. This finding aligns with existing research on knowledge internalization in LLMs~\cite{liu2022relationalmemoryaugmentedlanguage}.

\begin{table}[htbp]
\centering
\scriptsize
\begin{tabular}{@{}l l c c c c@{}}
\toprule
\textbf{Task} & \textbf{Setting} & \textbf{Metric} & \textbf{Normal} & \textbf{Individual} & \textbf{Ours} \\
\midrule
Effectiveness & Clean Model & FSR ($\uparrow$) & 68\% & 62\% & 100\% \\
\cmidrule(lr){1-6}
\multirow{3}{*}{Fine-tuning} 
& Alpaca & FSR ($\uparrow$) & 60\% & 48\% & 69\% \\
& ShareGPT & FSR ($\uparrow$) & 58\% & 44\% & 98\% \\
& Dolly & FSR ($\uparrow$) & 56\% & 34\% & 100\% \\
\cmidrule(lr){1-6}
\multirow{2}{*}{Perturbation} 
& Remove 5\% & FSR ($\uparrow$) & 67\% & 24\% & 100\% \\
& Remove 10\% & FSR ($\uparrow$) & 67\% & 24\% & 100\% \\
\cmidrule(lr){1-6}
Harmlessness & Task Perf. & ACC ($\uparrow$) & 0.559 & 0.560 & 0.588 \\
\bottomrule
\end{tabular}
\caption{Performance of Normal and Individual datasets on LLaMA2}
\label{tab:Ablation}
\end{table}

\section{Conclusion}

Our method of fingerprint model is to establish a close relationship between fingerprint information and make it firmly embedded in the model by establishing a family. Then, by extracting part of the prefix information of the fingerprint, the opponent model generates a long text, and compares it with our fingerprint text to determine whether there is a suspect of stealing our model. In the experiments, we also simulate a large number of real attack scenarios, and also evaluate the effectiveness, stealthiness, robustness and harmlessness by AUC scores. Compared with other fingerprinting methods, our KinGuard has a significant improvement in these performance. In the future, we also plan to embed the KinGuard into more models for testing to ensure its generality.

\footnotesize
\bibliographystyle{IEEEbib}
\bibliography{strings,refs}

\end{document}